\documentstyle[12pt]{article}

\title{Canonical reduction of two-dimensional gravity \\
for Particle Dynamics }          
\author{Tadayuki Ohta \\ 
Department of Physics, Miyagi University of Education \\
Aoba-Aramaki, Sendai 980, Japan 
\and 
Robert Mann\\ 
Department of Physics, University of Waterloo\\
Waterloo, Ontario, Canada N2L 3G1}  

\date{}

\begin{document}

\maketitle

\begin{abstract}
We develop the formalism for canonical reduction of 
$(1+1)$--dimensional gravity 
coupled with a set of point particles by eliminating constraints and 
imposing coordinate conditions. The formalism itself is quite analogous to 
the $(3+1)$--dimensional case; however in $(1+1)$ dimensions an auxiliary 
scalar field is shown to have an important role. The reduced Hamiltonian is 
expressed as a form of spatial integral of the second derivative of the 
scalar field. Since in $(1+1)$ dimensions there exists no dynamical 
degree of freedom of the gravitational field 
({\it i.e.} the transverse-traceless part of the metric tensor is zero), 
the reduced Hamiltonian is completely determined in terms of the
particles' canonical variables (coordinates and momenta). The explicit form 
of the Hamiltonian is calculated both in  post-linear and  post-Newtonian 
approximations.  
\end{abstract}

\section{Introduction}

Early versions of $(1+1)$--dimensional gravity \cite{r1,r2} 
have in recent years led to intensive study of a wide variety of
such theories, in large part because problems in
quantum gravity become much more mathematically tractable in this
context \cite{HS92}. In contrast to $(3+1)$--dimensional gravity, 
the action integrals for these theories must 
incorporate some dynamics in the form of an auxiliary (or dilaton)
field  since the Einstein tensor is topologically trivial in two dimensions.
Although most of these 
theories can be written in a generic form \cite{BanksMann} 
whose general properties can then be studied \cite{LMK,2djol,jchan}, 
there is often much to be learned by focussing on specific models
within this general class \cite{RST}.

One such theory has been extensively studied in many respects, including 
gravitational collapse, black holes, cosmological solutions and
quantization \cite{r3,r4,r5,rtquant}. Referred to as $R=T$ theory,
the specific form
of the coupling of the auxiliary field $\Psi$ to gravity is chosen so that
it decouples from the classical field equations in such a way as to
ensure that the evolution of the gravitational field is determined only 
by the matter stress-energy (and reciprocally) \cite{r3,r4}. In
this manner $R=T$ theory captures in two spacetime dimensions 
the essence of classical general relativity (as opposed to 
classical scalar-tensor theories), and has $(1+1)$--dimensional
analogs of many of its properties \cite{r4,r5}.  Indeed, the theory can 
be understood as the $D\rightarrow 2$ limit of general relativity
\cite{2dross}.

One of the fundamental problems of (1+1)-dimensional gravity is its
relationship 
to Newtonian gravity. This is in general a problematic issue \cite{jchan}.
For a system of particles the Hamiltonian in Newtonian
gravity in two dimensions is  
\begin{equation}
H=\sum_{a}\frac{p^{2}_{a}}{2m_{a}}
+\pi G\sum_{a}\sum_{b}m_{a}m_{b}\mid z_{a}-z_{b}\mid 
\end{equation}
\noindent
where $m_{a}$, $z_{a}$ and $p_{a}$ are the rest mass, the coordinate and
the momentum of $a$-th particle, respectively, and $G$ is the 
gravitational constant. We note that
the $R=T$ theory has been shown to have a Newtonian limit
\cite{r3,r5}. However the dynamical role
of the auxiliary field has not yet been fully analysed, a task which
we investigate in this paper.

More generally we shall, in the context of $R=T$ theory,
formulate a general framework for deriving
a Hamiltonian for a system of particles, which coincides in a slow 
motion, weak field limit with
the Hamiltonian (1).  In addition to the properties stated above, an
advantage of the $R=T$ model is that
it needs only one auxiliary field. As with the ADM formalism in 
$(3+1)$--dimensional theory, we develop a canonical reduction 
by eliminating 
constraints and imposing coordinate conditions, which are quite analogous
to the ADM conditions. In its final form the reduced Hamiltonian of the 
system is given as a spatial integral of the second derivative of the 
auxiliary scalar field $\Psi$. Since in two dimensions there exists no 
transverse-traceless part of  the metric tensor (which are the
real dynamical variables of gravitational field), the 
scalar field $\Psi$ is given as a function of the dynamical variables 
($z_{a}$, $p_{a}$) of the particles by solving the constraint equations. 
Then the  Hamiltonian is completely determined in terms of 
the coordinates and momenta of the particles.

The outline of our paper is as follows. Details of the reduction process are
described in section 2, where much attention is devoted to the 
transformation of the total generator of the whole system and 
the choice of the coordinate conditions, from which the reduced 
Hamiltonian is defined. The consistency of the canonical reduction is 
proved in section 3. It guarantees that the canonical equations of motion 
given by the reduced Hamiltonian are identical with the original geodesic 
equations. The explicit form of the Hamiltonian is calculated in section 
4 in a post-linear approximation (a series expansion in the coupling 
constant $G$) and in section 5 in a post-Newtonian approximation. While 
the former approximation is appropriate for the analysis 
of fast motion, the latter is adequate for treating the slow motion, 
weak field case. Section 6 is reserved for concluding remarks. The role 
of the $\Psi$ field is discussed in the simplest case of a single static 
source. Finally, an appendix illustrates
the relation between the equations of motion in our coordinate
conditions and those in other conditions.

\section{Canonical formalism for particle dynamics}
\renewcommand{\thefootnote}{\fnsymbol{footnote}}

The action integral for the gravitational field coupled with point 
particles is \cite{r4}

\noindent

\begin{eqnarray}
I&=&\int dx^{2}\left[
\frac{1}{2\kappa}\sqrt{-g}
\left\{\Psi R+\frac{1}{2}g^{\mu\nu}\nabla_{\mu}\Psi\nabla_{\nu}\Psi\right\}
\right.
\nonumber \\
&&\makebox[2em]{}-\left.\sum_{a} m_{a}\int d\tau_{a}
\left\{-g_{\mu\nu}(x)\frac{dz^{\mu}_{a}}{d\tau_{a}}
\frac{dz^{\nu}_{a}}{d\tau_{a}}\right\}^{1/2}\delta^{2}(x-z_{a}(\tau_{a}))
\right] 
\end{eqnarray}
\noindent
where $\Psi$ is the  auxiliary scalar field. Here $g_{\mu\nu}$,
$g$, $R$ and $\tau_{a}$ are the metric tensor of spacetime, 
det$(g_{\mu\nu})$,
the Ricci scalar and the proper time of $a$-th particle, respectively, 
and $\kappa=8\pi G/c^4$. The symbol $\nabla_{\mu}$ denotes the covariant 
derivative associated with $g_{\mu\nu}$.  

The field equations derived from the
variations $\delta\Psi$ and $\delta g_{\mu\nu}$ are
\begin{equation}
R-g^{\mu\nu}\nabla_{\mu}\nabla_{\nu}\Psi=R-\frac{1}{\sqrt{-g}}\partial_{\mu}
(\sqrt{-g}g^{\mu\nu}\partial_{\nu}\Psi)=0 
\end{equation}
\begin{equation}\label{e4}
\frac{1}{2}\nabla_{\mu}\Psi\nabla_{\nu}\Psi
-\frac{1}{4}g_{\mu\nu}\nabla^{\lambda}\Psi\nabla_{\lambda}\Psi
+g_{\mu\nu}\nabla^{\lambda}\nabla_{\lambda}\Psi
-\nabla_{\mu}\nabla_{\nu}\Psi=\kappa T_{\mu\nu} 
\end{equation}
where 
\begin{eqnarray}
T_{\mu\nu}&=&-\frac{2}{\sqrt{-g}}\frac{\delta\cal L\mit_{M}}{\delta g^{\mu\nu}}
\nonumber \\
&=&\sum_{a} m_{a}\int d\tau_{a}\frac{1}{\sqrt{-g}}
g_{\mu\sigma}g_{\nu\rho}\frac{dz^{\sigma}_{a}}{d\tau_{a}}
\frac{dz^{\rho}_{a}}{d\tau_{a}}\delta^{2}(x-z_{a}(\tau_{a}))\;\;  ,
\end{eqnarray}
$\cal L\mit_{M}$ being the matter Lagrangian given by the second term
in the brackets on the right hand side of (2). 
The geodesic equation  derived from the variation $\delta z^{\mu}_{a}$ is
\begin{equation}\label{geo}
\frac{d}{d\tau_{a}}
\left\{g_{\mu\nu}(z_{a})\frac{dz^{\nu}_{a}}{d\tau_{a}}\right\}
-\frac{1}{2}g_{\nu\lambda,\mu}(z_{a})\frac{dz^{\nu}_{a}}{d\tau_{a}}
\frac{dz^{\lambda}_{a}}{d\tau_{a}}=0 \;\;.
\end{equation}
The trace of Eq.(4) is
\begin{equation}
\nabla^{\mu}\nabla_{\mu}\Psi=\kappa T^{\mu}_{\;\;\mu} 
\end{equation}
which yields 
\begin{equation}\label{RT}
R=\kappa T^{\mu}_{\;\;\mu} \;\;.
\end{equation}

Hence particle dynamics in $R=T$ theory
may be described in terms of the 
equations (6) and (8), both of which are independent of the scalar field. 
Particulate matter (in terms of the trace of its stress-energy tensor) 
generates spacetime  curvature via (\ref{RT}); the effects of
curvature act back upon matter via (\ref{geo}).

Note that all three components of the
metric tensor cannot be determined from (\ref{RT}), since it is
only one equation. The two extra degrees of freedom are related to
the choice of coordinates.
If the coordinate conditions are chosen to be independent of 
$\Psi$, equation 
(\ref{RT}) determines the metric tensor completely. 
More generally however, we need to know the scalar 
field $\Psi$, through which the metric tensor is determined -- it
is this field that guarantees conservation of the stress-energy
tensor via (\ref{e4}). So far little 
attention has been paid to the role of the scalar field ~$\Psi$~.

To derive a Hamiltonian for a system of particles, we shall utilize the
canonical formalism \cite{r6}. Writing
$\gamma=g_{11},  N_{0}= (-g^{00})^{-1/2}, N_{1}= g_{10}$, 
the action (2) transforms to 
\begin{equation}\label{e9}
I=\int dx^{2}\left\{\sum_{a}p_{a}\dot{z}_{a}\delta(x-z_{a}(x^{0}))
+\pi\dot{\gamma}+\Pi\dot{\Psi}+N_{0}R^{0}+N_{1}R^{1}\right\} 
\end{equation}
where $\pi$ and $\Pi$ are conjugate momenta to $\gamma$ and $\Psi$
respectively, and 
\begin{eqnarray}
R^{0}&=&-\kappa\sqrt{\gamma}\gamma\pi^{2}+2\kappa\sqrt{\gamma}\pi\Pi
+\frac{1}{4\kappa\sqrt{\gamma}}(\Psi^{\prime})^{2}
-\frac{1}{\kappa}\left(\frac{\Psi^{\prime}}{\sqrt{\gamma}}\right)^{\prime}
-\sum_{a}\sqrt{\frac{p^{2}_{a}}{\gamma}+m^{2}_{a}}\;
\delta(x-z_{a}(x^{0}))
\nonumber \\
\\
R^{1}&=&\frac{\gamma^{\prime}}{\gamma}\pi-\frac{1}{\gamma}\Pi\Psi^{\prime}
+2\pi^{\prime}
+\sum_{a}\frac{p_{a}}{\gamma}\delta(x-z_{a}(x^{0})) \;\;.
\end{eqnarray}
Here and in the following we denote $\partial_{0}$ by a symbol 
$(\;\dot{}\;)$ and $\partial_{1}$ by a symbol $(\;^{\prime}\;)$.

Taking variations $\delta\gamma, \delta\pi, \delta N_{0},
\delta N_{1}, \delta\Psi, \delta\Pi, \delta z_{a}$ and $\delta p_{a}$ 
of the action (\ref{e9}), we have
\begin{eqnarray}
\dot{\pi}&+&N_{0}\left\{\frac{3\kappa}{2}\sqrt{\gamma}\pi^{2}
-\frac{\kappa}{\sqrt{\gamma}}\pi\Pi
+\frac{1}{8\kappa\sqrt{\gamma}\gamma}(\Psi^{\prime})^{2}
-\sum_{a}\frac{p^{2}_{a}}{2\gamma^{2}\sqrt{\frac{p^{2}_{a}}{\gamma}
+m^{2}_{a}}}\;\delta(x-z_{a}(x^{0}))\right\}
\nonumber \\
&+&N_{1}\left\{-\frac{1}{\gamma^{2}}\Pi\Psi^{\prime}
+\frac{\pi^{\prime}}{\gamma}
+\sum_{a}\frac{p_{a}}{\gamma^{2}}\;\delta(x-z_{a}(x^{0}))\right\}
\nonumber \\
&+&N^{\prime}_{0}\frac{1}{2\kappa\sqrt{\gamma}\gamma}\Psi^{\prime}
+N^{\prime}_{1}\frac{\pi}{\gamma}=0 
\end{eqnarray}
\begin{equation}
\dot{\gamma}-N_{0}(2\kappa\sqrt{\gamma}\gamma\pi-2\kappa\sqrt{\gamma}\Pi)
+N_{1}\frac{\gamma^{\prime}}{\gamma}-2N^{\prime}_{1}=0 
\end{equation}
\begin{equation}
R^{0}=0 
\end{equation}
\begin{equation}
R^{1}=0 
\end{equation}
\begin{equation}
\dot{\Pi}+\partial_{1}(-\frac{1}{\gamma}N_{1}\Pi
+\frac{1}{2\kappa\sqrt{\gamma}}N_{0}\Psi^{\prime}
+\frac{1}{\kappa\sqrt{\gamma}}N^{\prime}_{0})=0 
\end{equation}
\begin{equation}
\dot{\Psi}+N_{0}(2\kappa\sqrt{\gamma}\pi)-N_{1}(\frac{1}{\gamma}
\Psi^{\prime})=0
\end{equation}
\begin{eqnarray}
\dot{p}_{a}&+&\frac{\partial N_{0}}{\partial z_{a}}\sqrt{\frac{p^{2}_{a}}
{\gamma}+m^{2}_{a}}-\frac{N_{0}}{2\sqrt{\frac{p^{2}_{a}}{\gamma}+m^{2}_{a}}}
\frac{p^{2}_{a}}{\gamma^{2}}\frac{\partial\gamma}{\partial z_{a}}
\nonumber \\
&-&\frac{\partial N_{1}}{\partial z_{a}}\frac{p_{a}}{\gamma}
+N_{1}\frac{p_{a}}{\gamma^{2}}\frac{\partial\gamma}{\partial z_{a}}=0 
\end{eqnarray}
\begin{equation}
\dot{z_{a}}-N_{0}\frac{\frac{p_{a}}{\gamma}}{\sqrt{\frac{p^{2}_{a}}{\gamma}
+m^{2}_{a}}}
+\frac{N_{1}}{\gamma}=0 \;\;.
\end{equation}
\noindent
In the equations (18) and (19), all metric components ($N_{0}$, $N_{1}$, 
$\gamma$) are evaluated at the point $x=z_{a}$ and
\[\frac{\partial f}{\partial z_{a}}\equiv 
\left.\frac{\partial f(x)}{\partial x}\right|_{x=z_{a}}
\]
It can be shown that this set of equations is equivalent to the equations
(3), (4) and (6). Here we show explicitly the equivalence of (18) and (19)
to the geodesic equation (6). We express the $(\mu=1)$ component of (6) 
in terms  of $N_{0}, N_{1}$ and $\gamma$ and their derivatives. 
Using the relations
\[d\tau_{a}=dt\left\{N^{2}_{0}-\frac{1}{\gamma}(N_{1}+\gamma\dot{z}_{a})^{2}
\right\}^{\frac{1}{2}}
\]
we get
\begin{eqnarray}
\frac{d}{dt}\left\{\frac{N_{1}+\gamma\dot{z}_{a}}{\left[N^{2}_{0}
-\frac{1}{\gamma}(N_{1}+\gamma\dot{z}_{a})^2\right]^{\frac{1}{2}}}\right\}
+\frac{1}{\left[N^{2}_{0}-\frac{1}{\gamma}(N_{1}+\gamma\dot{z}_{a})^2\right]
^{\frac{1}{2}}}\left\{N_{0}\frac{\partial N_{0}}{\partial z_{a}}
-\left(\frac{N_{1}}{\gamma}+\dot{z}_{a}\right)\frac{\partial N_{1}}
{\partial z_{a}}\right.
\nonumber \\
\makebox[10em]{}\left.+\frac{1}{2}\left[\frac{N^{2}_{1}}{\gamma^{2}}
-(\dot{z}_{a})^{2}\right]\frac{\partial\gamma}{\partial z_{a}}\right\}=0\;\;.
\end{eqnarray}
Eliminating $p_{a}$ from (18) and (19) yields an equation identical
with (20).

{}From the expression (9), the total generator (obtained from variations at
the end point) is given by
\begin{equation}\label{gen1}
G=\int dx\left\{\sum_{a}p_{a}\delta(x-z_{a})\delta z_{a}
+\pi\delta h-\Psi\delta\Pi\right\} \;\;\;,
\end{equation}
where the above form has been obtained by adding a total time 
derivative $-\partial_{0}(\Pi\Psi)$ to the original action (9), 
and where the  constraint equations (14) and (15) have been
taken into account. In the above $h=1+\gamma$; the advantage of using 
$h$ will become clear later \cite{r7,r8}.

We must now identify the dynamic and gauge character of the variables
which appear in the generator (\ref{gen1}).  By considering the 
constraint equations (14) and (15), we see that the only linear
terms there are $\frac{\Psi^{\prime}}{\sqrt{h-1}}$ and $\pi^{\prime}$
respectively.  The equations may therefore be solved for these quantities
in terms of the dynamical and gauge ({\it i.e.} co-ordinate) degrees of
freedom. By writing $\Pi = \frac{1}{\triangle}\Pi^{\prime\prime}$, 
where $1/\triangle$ is the inverse of the operator 
$\triangle=\partial^{2}/\partial x^{2}$ 
with appropriate boundary condition, we find that
the generator becomes
\begin{eqnarray}
G&=&\int dx\left\{\sum_{a}p_{a}\delta(x-z_{a})\delta z_{a}
-\left[-\frac{1}{\kappa}\left(\frac{\Psi^{\prime}}{\sqrt{h-1}}
\right)^{\prime}\right]
\delta\left[-\frac{\kappa}{\triangle}\left(\sqrt{h-1}\frac{1}{\triangle}
\Pi^{\prime}\right)^{\prime}\right]
\right.
\nonumber \\
&&\makebox[5em]{}\left.-\left[2\pi^{\prime}-\left(\frac{\Psi^{\prime}}{h-1}
\right)^{\prime}\frac{1}{\triangle}\Pi^{\prime}-\frac{1}{h-1}\Pi\Psi^{\prime}
\right]\delta\left(\frac{1}{2\triangle}h^{\prime}\right)\right\} \;\;.
\label{e22}
\end{eqnarray}
where in obtaining this form, we have discarded surface terms.

The form (\ref{e22}) of the generator is analogous what is obtained
in $(3+1)$ dimensions under an orthogonal decomposition of
the hypersurface metric and its conjugate momentum \cite{r7,r8}.
We therefore propose adopting the coordinate conditions
\begin{eqnarray}
x&=&\frac{1}{2\triangle}h^{\prime} 
\\
t&=&-\frac{\kappa}{\triangle}\left(\sqrt{h-1}\frac{1}{\triangle}\Pi^{\prime}
\right)^{\prime} \;\;
\end{eqnarray}
which will then allow us to identify the Hamiltonian and momentum densities
as the respective coefficients of $\delta t$ and $\delta x$ in
the canonical form of the generator. Taking a spatial derivative
of these coordinate conditions yields their
differential form, which does not explicitly depend upon the coordinates
$(t,x)$. This leads to the choices
\begin{eqnarray}
h=2 &\longrightarrow& \gamma=1 
\\
\frac{1}{\triangle}\Pi^{\prime}=0 &\longrightarrow& \Pi=0 \;\; 
\end{eqnarray}
which may then be inserted in equations (12) and (19) to
solve for the relevant physical degrees of freedom.

Specifically, we may insert these choices into (14) and (15) to
solve for the Hamiltonian and momentum densities. We then find
that the generator (\ref{e22}) may be expressed in the canonical form
\begin{equation}
G=\int dx\left\{\sum_{a}p_{a}\delta(x-z_{a})\delta z_{a}
-\cal T\mit_{0 \mu}\delta x^{\mu}\right\} 
\end{equation}
where
\begin{eqnarray}
\cal T\mit_{0 0}&=& \cal H\mit =-\frac{1}{\kappa}\triangle\rm\Psi\mit
\\
\cal T\mit_{0 1}&=&2\pi^{\prime} \;\;.
\end{eqnarray}
Thus, $\cal H\mit$ is the Hamiltonian density of the system 
and $\cal T\mit_{0 1}$ is the momentum density. Note that
since the constraints (14) and (15) have already been imposed, 
${\cal T}_{0\mu}$ is expressed in terms of the canonical variables $z_{a}$
and $p_{a}$ as ${\cal T}_{0\mu}(x, z_{a}, p_{a})$, by solving these
constraints. With the  coordinate choices (25) and (26), they lead to
\begin{equation}
\triangle\Psi-\frac{1}{4}(\Psi^{\prime})^{2}
+\kappa^{2}\pi^{2}+\kappa\sum_{a}\sqrt{p^{2}_{a}+m^{2}_{a}}
\delta(x-z_{a})=0 \;\;,
\end{equation}
\begin{equation}
2\pi^{\prime}+\sum_{a}p_{a}\delta(x-z_{a})=0 \;\;.
\end{equation}

Following a similar procedure as that used to transform the generator $G$,
we rewrite the action integral (9) as
\begin{eqnarray}
I&=&\int d^{2}x\left\{\sum_{a}p_{a}\dot{z}_{a}\delta(x-z_{a})
-\left[-\frac{1}{\kappa}\left(\frac{\Psi^{\prime}}{\sqrt{h-1}}\right)^{\prime}
\right]\partial_{0}\left[-\frac{\kappa}{\triangle}\left(\sqrt{h-1}
\frac{1}{\triangle}\Pi^{\prime}\right)^{\prime}\right]
\right.
\nonumber \\
&&\makebox[2em]{}\left.
-\left[2\pi^{\prime}-\left(\frac{\Psi^{\prime}}{h-1}\right)^{\prime}
\frac{1}{\triangle}\Pi^{\prime}-\frac{1}{h-1}\Pi\Psi^{\prime}\right]
\partial_{0}\left(\frac{1}{2\triangle}h^{\prime}\right)+N_{\mu}R^{\mu}
\right\}\label{e32}
\end{eqnarray}
where we have discarded surface terms.
Eliminating the constraints (14) and (15) and imposing the coordinate
conditions (23) and (24), the action integral reduces to
\begin{equation}
I_{R}=\int dx^{2}\left\{\sum_{a}p_{a}\dot{z}_{a}\delta(x-z_{a})
-\cal H\mit\right\}\;\;.
\end{equation}

\noindent
Thus the reduced Hamiltonian for the system of particles is 
\begin{equation}
H=\int dx \cal H\mit =-\frac{1}{\kappa}\int dx \triangle\Psi 
\end{equation}
where $\Psi$ is a function of $z_{a}$ and $p_{a}$ and is determined
by solving the constraint equations (30) and (31).
This expression is analogous to the reduced Hamiltonian in 
$(3+1)$ dimensional general relativity \cite{r6,r7,r8}.

We pause to comment on the relationship between the integral 
forms (23) and (24) of the coordinate conditions and the 
differential forms (25) and (26). These two forms are equivalent 
only when one retains the appropriate boundary conditions for the integral 
operator $1/\triangle$. A proper treatment \cite{r9,r10} entails the 
insertion of a regulator  $\exp(-\alpha |x|)$ in the left-hand sides 
of (23) and (24), yielding
$$
{g} :=   2\,{\rm e}^{(\, - { \alpha}\,{|x|}\,)}\,(\, 1 - 
{ \alpha}\,{|x|}\,) - 1
$$
and 
$$
{ \Pi} := \frac{d}{dx}\left({\displaystyle \frac {{-\alpha}\,{t}\,{\rm e}
^{(\, - { \alpha}\,{|x|}\,)}{\mbox{sgn}}(x)}{(\, 2\,{\rm e}^{(\, - { \alpha}\,{|x|}\,
)} - 2\,{|x|}\,{ \alpha}\,{\rm e}^{(\, - { \alpha}\,{|x|}\,)} - 1\,)
^{1/2}\,{ \kappa}}}\right)
$$
in place of (25) and (26) respectively. We must then insert these
quantities in (12) -- (19),
taking the limit $\alpha \to 0$ at the end of the calculation
(implicitly assuming $\alpha|x| < 1$).
This turns out to be equivalent to inserting (25) and (26) in
these equations.  The action (33) is recovered by a similar limiting
procedure.

This situation is analogous with that in (3+1) dimensions
dimensions as discussed by ADM in \cite{r10}
(cf. Eqs. (80), (81), (84) and  (85) in section 6).  
We shall discuss these points further in section 6. Note that
the coordinate conditions in integral form are needed only at the stage of 
transforming the generator and defining the Hamiltonian density. After 
fixing the formalism, we need only the differential forms to solve the 
constraints and to evaluate the Hamiltonian, a problem which is
treated in the  following sections.

\section{Consistency of the canonical reduction}

Following a procedure similar to the original ADM argument 
\cite{r7,r9}, we shall demonstrate here that the canonical equations of 
motion derived from the reduced Hamiltonian (34) are identical with 
equations (18) and (19), which are in turn equivalent to the original 
geodesic equation (6).

We start from the action integral (9). The variations with respect to 
$p_{a}$ and $z_{a}$ lead to the equations of motion 
\begin{eqnarray}
\dot{z}_{a}=-\int d^{2}y N_{\mu}(y)\frac{\delta R^{\mu}(y)}{\delta p_{a}(t)}
\\
\dot{p}_{a}=\int d^{2}y N_{\mu}(y)\frac{\delta R^{\mu}(y)}
{\delta z_{a}(t)}\;\;.
\end{eqnarray}
These equations are identical with equations (18) and (19). Our 
purpose is to prove that these equations lead, when they are combined 
with the constraints (14) and (15), to the equations in the reduced 
formalism. 

Defining
\begin{eqnarray}
P_{0}(x)&\equiv&-\frac{1}{\kappa}\triangle\Psi(x)
\\
P_{1}(x)&\equiv&2\pi^{\prime}(x)
\end{eqnarray}
and imposing the coordinate conditions, the action (\ref{e32})
may be expressed as
\begin{equation}
I=\int d^{2}x\left\{\sum_{a}p_{a}\dot{z}_{a}\delta(x-z_{a})-P_{0}
+N_{\mu}R^{\mu}\right\}\;\;.
\end{equation}
where we have not yet imposed the constraints $R^{\mu}=0$.
By solving the constraint equations, we get
\begin{eqnarray}
P_{0}(x)&=&{\cal T}_{0 0}(x,z_{a},p_{a})\equiv \cal H\mit
\\
P_{1}(x)&=&{\cal T}_{0 1}(x,z_{a},p_{a})\;\;
\end{eqnarray}
where ${\cal T}_{0 \mu}(x, z_{a}, p_{a})$ was 
described in the previous section.

Expanding $R^{\mu}$ in a functional Taylor series about the point 
$P_{\mu}={\cal T}_{0 \mu}$ gives
\begin{equation}
R^{\mu}(x)=\int d^{2}y\left[P_{\nu}(y)-{\cal T}_{0 \nu}
(y, z_{a}, p_{a})\right]
\left[\frac{\delta R^{\mu}(x)}{\delta P_{\nu}(y)}\right]_{P=\cal T\mit}
+\cdot\cdot\cdot\;\;.
\end{equation}
where the coordinate conditions have already been imposed.
Substituting this expansion into the right hand side of (39)
and taking the variation of the action $I$ with respect to $P^{\nu}$,
yields
\begin{equation}
-\delta_{\nu 0}+\int d^{2}y N_{\mu}(y)\left\{\left[\frac{\delta R^{\mu}(y)}
{\delta P_{\nu}(x)}\right]_{P=\cal T\mit}+\cdot\cdot\cdot\right\}
=0
\end{equation}
where the terms represented by $\cdot\cdot\cdot$ contain 
$[P_{\nu}-{\cal T}_{0 \nu}]$ as a factor.

Now we need the relation which is valid after imposing the constraints. Then,
by requiring $R^{\mu}=0$, we have
\begin{equation}
-\delta_{\nu 0}+\int d^{2}y N_{\mu}(y)\left[\frac{\delta R^{\mu}(y)}
{\delta P_{\nu}(x)}\right]_{P=\cal T}=0\;\;.
\end{equation}
Insertion of (42) into (35) leads to
\begin{eqnarray}
\dot{z}_{a}&=&-\int d^{2}y N_{\mu}(y)\int d^{2}x\left\{
-\frac{\partial {\cal T}_{0 \nu}}
{\partial p_{a}(t)}\left[\frac{\delta R^{\mu}(y)}{\delta P_{\nu}(x)}\right]
_{P=\cal T}\right.
\nonumber \\
&&\makebox[5em]{}\left.+(P_{\nu}(x)-{\cal T}_{0 \nu}(x,z_{a},p_{a}))
\frac{\partial}{\partial p_{a}(x)}\left[\frac{\delta R^{\mu}(y)}{\delta P_{\nu}
(x)}\right]_{P=\cal T\mit}+\cdot\cdot\cdot\right\}
\end{eqnarray}
Imposing the constraint equations $R^{\mu}=0$ yields for (45) 
\begin{equation}
\dot{z}_{a}
=\int d^{2}x\frac{\partial {\cal T}_{0 \nu}}{\partial p_{a}(t)}
\int d^{2}y N_{\mu}(y)\left[\frac{\delta R^{\mu}(y)}{\delta P_{\nu}(x)}
\right]_{P=\cal T}
\end{equation}
As a consequence of the relation (44), (46) becomes 
\begin{eqnarray}
\dot{z}_{a}&=&\int d^{2}x\frac{\partial {\cal T}_{0 0}}
{\partial p_{a}}
\nonumber \\
&=&\frac{\partial H}{\partial p_{a}}
\end{eqnarray}

In the same way we can show that the equations of motion (36) is identical 
with
\begin{equation}
\dot{p}_{a}=-\frac{\partial H}{\partial z_{a}}
\end{equation}
Thus the consistency of the reduced canonical formalism is proved.

\section{Post-linear approximation}

For a direct calculation of the reduced Hamiltonian (34), we have to solve
the constraint equations (30) and (31). However, it seems quite a difficult
task to get an exact solution except in the case of a single static source.
So we need some approximation method. 

In this section we apply an iterative method successively to the integrand
of the expression (34) with the use of the equations (30) and (31). This
provides us with an expansion of (34) in powers of $\kappa$, which we
refer to as the post-linear approximation.

First, substituting $\triangle\Psi$ given by the equation (30) into the 
integrand and performing a partial integration, we have
\begin{eqnarray}
H&=&\int dx\left\{-\frac{1}{4\kappa}\left(\Psi^{\prime}\right)^{2}
+\kappa\left(\chi^{\prime}\right)^{2}+\sum_{a}\sqrt{p^{2}_{a}+m^{2}_{a}}
\;\delta(x-z_{a})\right\}
\nonumber \\
&=&\sum_{a}\sqrt{p^{2}_{a}+m^{2}_{a}}+\int dx\left\{
\frac{1}{4\kappa}\Psi\triangle\Psi-\kappa\chi\triangle\chi\right\}+S_{1}
\end{eqnarray}
where $\chi$ is defined by $\chi^{\prime}\equiv\pi$ and
\begin{equation}
S_{1}=\left[-\frac{1}{4\kappa}\Psi\Psi^{\prime}+\kappa\chi\chi^{\prime}
\right]^{\infty}_{-\infty}\;\;.
\end{equation}
Noting that $\Psi$ is of order $\kappa$, we
substitute (30) for $\triangle\Psi$ and (31) for
$\triangle\chi$ into the integrand of the second term on the right hand 
side of (49). Iterating twice in terms of $\kappa$ yields 
the expression
\begin{eqnarray}
H&=&\sum_{a}\sqrt{p^{2}_{a}+m^{2}_{a}}\left\{1-\frac{1}{4}\Psi(z_{a})
+\frac{1}{32}\Psi(z_{a})\Psi(z_{a})\right\}
\nonumber \\
&&\makebox[2em]{}+\frac{\kappa}{2}\sum_{a}p_{a}\chi(z_{a})
\left\{1-\frac{1}{4}\Psi(z_{a})\right\}+\frac{\kappa^{2}}{8}\sum_{a}
\sqrt{p^{2}_{a}+m^{2}_{a}}\;\chi(z_{a})\chi(z_{a})+S_{2}
\nonumber \\
&&\makebox[2em]{}+\int dx\left\{\frac{\kappa^{3}}{8}\chi^{2}
\left(\chi^{\prime}\right)^{2}
-\frac{\kappa}{32}\chi^{2}\left(\Psi^{\prime}\right)^{2}
+\frac{\kappa}{32}\Psi^{2}\left(\chi^{\prime}\right)^{2}
-\frac{1}{128\kappa}\Psi^{2}\left(\Psi^{\prime}\right)^{2}\right\}
\label{e51}
\end{eqnarray}
where 
\begin{equation}\label{e52}
S_{2}=\left[-\frac{1}{4\kappa}\Psi\Psi^{\prime}+\kappa\chi\chi^{\prime}
-\frac{\kappa}{8}\left\{\Psi\left(\chi^{2}\right)^{\prime}
-\Psi^{\prime}\chi^{2}\right\}+\frac{1}{32\kappa}\Psi^{2}\Psi^{\prime}
\right]^{\infty}_{-\infty}\;\;.
\end{equation}
The last term in (\ref{e51}) contributes at
order $\kappa^{3}$ . This iteration can be continued 
successively and corresponds to a series expansion in $\kappa$. 

In contrast to the $(3+1)$--dimensional situation, 
we encounter a subtle problem 
when we try to extract an explicit form of the Hamiltonian under
this iteration scheme. In $(1+1)$ dimensions the surface terms such 
as $S_{1}$ and $S_{2}$ arising 
in the process of the calculation do not necessarily vanish and make the 
Hamiltonian indefinite. This is because the dimensionless potential 
$Gmr/c^{2}$ becomes infinite at spatial infinity, in contrast to
$(3+1)$ dimensions where the corresponding quantity $Gm/rc^{2}$ 
vanishes at spatial infinity thereby assuring that the 
associated surface terms vanish. 
It is therefore an important task to find a solution of 
$\Psi$ and $\chi$ which makes the surface term vanish at a given
order in $\kappa$. This is related to the choice of
boundary condition.

{}From the expressions of $S_{1}$ and $S_{2}$, we can infer a simple choice
of the boundary condition as follows. 

\noindent
\hspace*{1cm}For $f(x)\equiv\Psi^{2}-4\kappa^{2}\chi^{2}$, 
\begin{equation}
\makebox[2em]{}f(x)=0 \makebox[1em]{}\mbox{and}\makebox[1em]{}
f^{\prime}(x)=0 \makebox[2em]{}\mbox{in a region}\makebox[1em]{} 
\mid x\mid>>\mid z_{a}\mid \makebox[1em]{} \mbox{for all}
\makebox[0.5em]{} a.
\end{equation}

\vspace{3mm}
\noindent
It is easily checked that under this boundary condition, the surface terms
$S_{1}$ and $S_{2}$ exactly vanish, since they are proportional to
$f^\prime$ and $(4-\Psi)f^\prime + f\Psi^\prime$ respectively.

Now let us try to obtain the Hamiltonian up to $\kappa^{2}$ (the 
2nd-post-linear approximation). 
First we expand $\Psi$ and $\chi$ in a power series in $\kappa$ 
\begin{eqnarray}
\Psi&=&\kappa\Psi^{(1)}+\kappa^{2}\Psi^{(2)}+\cdot\cdot\cdot 
\\
\chi&=&\chi^{(0)}+\kappa\chi^{(1)}+\cdot\cdot\cdot\;\;. 
\end{eqnarray}
Substituting this expansion into the equations (30) and (31), we get 
\begin{eqnarray}
\triangle\Psi^{(1)}&=&-\sum_{a}\sqrt{p^{2}_{a}+m^{2}_{a}}\;\delta(x-z_{a})
\\
\triangle\Psi^{(2)}&=&-\left(\chi^{(0)\prime}\right)^{2}
+\frac{1}{4}\left(\Psi^{(1)\prime}\right)^{2}
\\
\triangle\chi^{(0)}&=&-\frac{1}{2}\sum_{a}p_{a}\delta(x-z_{a})
\\
\triangle\chi^{(1)}&=&0\;\;.
\end{eqnarray}

The solutions which satisfy the boundary condition (53) are
\begin{eqnarray}
\Psi^{(1)}&=&-\frac{1}{2}\sum_{a}\left\{\sqrt{p^{2}_{a}+m^{2}_{a}}\;r_{a}
+\epsilon\;p_{a}(x-z_{a})\right\}
\\
\chi^{(0)}&=&-\frac{1}{4}\sum_{a}\left\{p_{a}r_{a}
+\epsilon\;\sqrt{p^{2}_{a}+m^{2}_{a}}\;(x-z_{a})\right\}
\end{eqnarray}

\begin{eqnarray}
\Psi^{(2)}&=&-\frac{1}{2}\left(\frac{1}{4}\right)^{2}\left\{\left[
\sum_{a}p_{a}r_{a}+\epsilon\;\sum_{a}\sqrt{p^{2}_{a}+m^{2}_{a}}\;
(x-z_{a})\right]^{2}
-\left[\sum_{a}\sqrt{p^{2}_{a}+m^{2}_{a}}\;r_{a}
+\epsilon\;\sum_{a}p_{a}(x-z_{a})\right]^{2}\right\}
\nonumber \\
&&-\left(\frac{1}{4}\right)^{2}\left\{\sum_{a}\sqrt{p^{2}_{a}+m^{2}_{a}}
\;r_{a}
\left[\sum_{b}\sqrt{p^{2}_{b}+m^{2}_{b}}\;r_{ab}
+\epsilon\;\sum_{b}p_{b}(z_{a}-z_{b})\right]
\right.
\nonumber \\
&&\makebox[4em]{}\left.-\sum_{a}p_{a}r_{a}\left[\sum_{b}p_{b}r_{ab}
+\epsilon\;\sum_{b}\sqrt{p^{2}_{b}+m^{2}_{b}}\;(z_{a}-z_{b})\right]\right\}
\\
\chi^{(1)}&=&-\frac{\epsilon}{2}\left(\frac{1}{4}\right)^{2}\left\{
\sum_{a}\sqrt{p^{2}_{a}+m^{2}_{a}}\;(x-z_{a})
\left[\sum_{b}\sqrt{p^{2}_{b}+m^{2}_{b}}\;r_{ab}
+\epsilon\;\sum_{b}p_{b}(z_{a}-z_{b})\right]\right.
\nonumber \\
&&\makebox[4em]{}\left.-\sum_{a}p_{a}(x-z_{a})\left[\sum_{b}p_{b}r_{ab}
+\epsilon\;\sum_{b}\sqrt{p^{2}_{b}+m^{2}_{b}}\;(z_{a}-z_{b})\right]
\right\}\;\;.
\end{eqnarray}
where $r_a \equiv |x-z_a|$.  

In these solutions we introduced a constant of integration $\epsilon$,
satifying $\epsilon^{2}=1$. We have two types of  solutions
corresponding to $\epsilon=1$ and $\epsilon=-1$, which are related to
each other under time reversal. This $\epsilon$ factor guarantees the 
invariance of the whole theory under the time reversal.

The boundary condition is checked as follows. Up to order $\kappa^{3}$
we have 
\begin{eqnarray}
\lefteqn{\Psi^{2}-4\kappa^{2}\chi^{2}}
\nonumber \\
&=&\frac{\kappa^{2}}{4}\sum_{a}\sum_{b}
\left\{\left[\sqrt{p^{2}_{a}+m^{2}_{a}}\sqrt{p^{2}_{b}+m^{2}_{b}}-p_{a}p_{b}
\right]\left(1-\frac{\kappa}{8}\sum_{c}\left[\sqrt{p^{2}_{c}+m^{2}_{c}}\;r_{c}
+\epsilon\;p_{c}(x-z_{c})\right]\right)\right.
\nonumber \\
&&\makebox[4em]{}\left.+\frac{\kappa}{4}\sqrt{p^{2}_{a}+m^{2}_{a}}
\sum_{c}\left[\sqrt{p^{2}_{b}+m^{2}_{b}}\;A_{bc}-p_{b}B_{bc}\right]\right\}
[r_{a}r_{b}-(x-z_{a})(x-z_{b})]
\nonumber \\
&&+\epsilon\;\frac{\kappa^{2}}{2}\sum_{a}\sum_{b}
\left\{\sqrt{p^{2}_{a}+m^{2}_{a}}\;p_{b}
\left(1-\frac{\kappa}{8}\sum_{c}\left[\sqrt{p^{2}_{c}+m^{2}_{c}}\;r_{c}
+\epsilon\;p_{c}(x-z_{c})\right]\right)\right.
\nonumber \\
&&\makebox[4em]{}\left.-\frac{\kappa}{8}p_{a}\sum_{c}
\left[\sqrt{p^{2}_{b}+m^{2}_{b}}\;A_{bc}-p_{b}B_{bc}\right]\right\}
[r_{a}(x-z_{b})-(x-z_{a})r_{b}]
\nonumber
\end{eqnarray}
where
\begin{eqnarray}
A_{bc}&=&\sqrt{p^{2}_{c}+m^{2}_{c}}\;r_{bc}+\epsilon\;p_{c}(z_{b}-z_{c})
\nonumber \\
B_{bc}&=&p_{c}r_{bc}+\epsilon\;\sqrt{p^{2}_{c}+m^{2}_{c}}(z_{b}-z_{c})\;\;.
\nonumber 
\end{eqnarray}
Since both $[r_{a}r_{b}-(x-z_{a})(x-z_{b})]$ and 
$[r_{a}(x-z_{b})-(x-z_{a})r_{b}]$ vanish in a region 
$\mid x\mid >> \mid z_{a}\mid, \mid z_{b}\mid$, 
the boundary condition is satisfied.

The 2nd-post-linear Hamiltonian is therefore unambiguously determined 
to be
\begin{eqnarray}
H&=&\sum_{a}\sqrt{p^{2}_{a}+m^{2}_{a}}\left\{1-\frac{1}{4}\Psi(z_{a})
+\frac{1}{32}\Psi(z_{a})\Psi(z_{a})\right\}
\nonumber \\
&&\makebox[2em]{}+\frac{\kappa}{2}\sum_{a}p_{a}\chi(z_{a})
\left\{1-\frac{1}{4}\Psi(z_{a})\right\}
+\frac{\kappa^{2}}{8}\sum_{a}\sqrt{p^{2}_{a}+m^{2}_{a}}\;\chi(z_{a})
\chi(z_{a})
\nonumber \\
&=&\sum_{a}\sqrt{p^{2}_{a}+m^{2}_{a}}
+\frac{\kappa}{8}\sum_{a}\sum_{b}\left(\sqrt{p^{2}_{a}+m^{2}_{a}}
\sqrt{p^{2}_{b}+m^{2}_{b}}-p_{a}p_{b}\right)r_{ab}
\nonumber \\
&&+\frac{\epsilon\kappa}{8}\sum_{a}\sum_{b}
\left(\sqrt{p^{2}_{a}+m^{2}_{a}}\;p_{b}
-p_{a}\sqrt{p^{2}_{b}+m^{2}_{b}}\right)(z_{a}-z_{b})
\nonumber \\
&&+\frac{1}{4}\left(\frac{\kappa}{4}\right)^{2}
\left\{\sum_{a}\sqrt{p^{2}_{a}+m^{2}_{a}}
\left[\sum_{b}p_{b}r_{ab}
+\epsilon\;\sum_{b}\sqrt{p^{2}_{b}+m^{2}_{b}}(z_{a}-z_{b})\right]^{2}\right.
\nonumber \\
&&\left.-\sum_{a}p_{a}\left[\sum_{b}p_{b}r_{ab}
+\epsilon\;\sum_{b}\sqrt{p^{2}_{b}+m^{2}_{b}}(z_{a}-z_{b})\right]
\left[\sum_{c}\sqrt{p^{2}_{c}+m^{2}_{c}}\;r_{ac}
+\epsilon\;\sum_{c}p_{c}(z_{a}-z_{c})\right]\right.
\nonumber \\
&&\left.+\sum_{a}\sum_{b}\left[\sqrt{p^{2}_{a}+m^{2}_{a}}\sqrt{p^{2}_{b}
+m^{2}_{b}}\;r_{ab}
-\epsilon\;p_{a}\sqrt{p^{2}_{b}+m^{2}_{b}}(z_{a}-z_{b})\right]\right.
\nonumber \\
&&\makebox[8em]{}\left.\times\left[\sum_{c}\sqrt{p^{2}_{c}+m^{2}_{c}}\;r_{bc}
+\epsilon\;\sum_{c}p_{c}(z_{b}-z_{c})
\right]\right.
\nonumber \\
&&\left.-\sum_{a}\sum_{b}\left[\sqrt{p^{2}_{a}+m^{2}_{a}}\;p_{b}r_{ab}
-\epsilon\;p_{a}p_{b}(z_{a}-z_{b})\right]\left[\sum_{c}p_{c}r_{bc}
+\epsilon\;\sum_{c}\sqrt{p^{2}_{c}+m^{2}_{c}}(z_{b}-z_{c})\right]\right\}\;\;.
\nonumber \\
\end{eqnarray}

As shown in the above development, the $\kappa$-expansion is the successive 
approximation in the background of Minkowskian space-time and the terms in 
each order of $\kappa$ preserve relativistic forms. This approximation
is therfore appropriate for describing relativistic fast-motion 
of the particles and can be carried out to any desired
order of $\kappa$.

\section{Post-Newtonian Hamiltonian and the redefinition of canonical 
variables}

To compare $R=T$ theory with Newtonian gravity, we 
need an approximation method applicable to a slow motion in a weak field.
It is provided by the so-called $c^{-1}$ expansion. All terms appropriate 
to the post-Newtonian approximation (up to the order of $c^{-4}$) are 
included in the post-linear Hamiltonian (64). Noting 
that both $p_a^2/m_a^2$ and $\sqrt{\kappa}$ are of the order of $c^{-2}$,
we find from (64)
\begin{eqnarray}
H&=&\sum_{a}m_{a}+\sum_{a}\frac{p^{2}_{a}}{2m_{a}}+\frac{\kappa}{8}\sum_{a}
\sum_{b}m_{a}m_{b}r_{ab}
+\frac{\epsilon\kappa}{8}\sum_{a}\sum_{b}(m_{a}p_{b}-m_{b}p_{a})(z_{a}-z_{b})
\nonumber \\
&&-\sum_{a}\frac{p^{4}_{a}}{8m^{3}_{a}}+\frac{\kappa}{8}\sum_{a}\sum_{b}
m_{a}\frac{p^{2}_{b}}{m_{b}}r_{ab}-\frac{\kappa}{8}\sum_{a}\sum_{b}p_{a}p_{b}
r_{ab}
\nonumber \\
&&+\frac{1}{4}\left(\frac{\kappa}{4}\right)^{2}\sum_{a}\sum_{b}\sum_{c}
m_{a}m_{b}m_{c}r_{ab}r_{ac}
+\frac{1}{4}\left(\frac{\kappa}{4}\right)^{2}\sum_{a}\sum_{b}\sum_{c}
m_{a}m_{b}m_{c}(z_{a}-z_{b})(z_{a}-z_{c})\;\;.
\nonumber \\
\end{eqnarray}
The corresponding solutions of $\Psi$ and $\chi$ are 
\begin{eqnarray}
\Psi&=&-\frac{\kappa}{2}\sum_{a}m_{a}r_{a}
-\frac{\epsilon\kappa}{2}\sum_{a}p_{a}(x-z_{a})
-\frac{\kappa}{4}\sum_{a}\frac{p^{2}_{a}}{m_{a}}r_{a}
+\frac{1}{2}\left(\frac{\kappa}{4}\right)^{2}\left(\sum_{a}m_{a}r_{a}\right)
^{2}
\nonumber \\
&&-\frac{1}{2}\left(\frac{\kappa}{4}\right)^{2}\left(\sum_{a}m_{a}(x-z_{a})
\right)^{2}-\left(\frac{\kappa}{4}\right)^{2}\sum_{a}\sum_{b}m_{a}m_{b}
r_{a}r_{ab}
\\
\chi&=&-\frac{\epsilon}{4}\sum_{a}m_{a}(x-z_{a})
-\frac{1}{4}\sum_{a}p_{a}r_{a}\;\;.
\end{eqnarray}

The canonical equations of motion yield
\begin{eqnarray}
\dot{z}_{a}&=&\frac{\partial H}{\partial p_{a}}
\nonumber \\
&=&\frac{p_{a}}{m_{a}}-\frac{\epsilon\kappa}{4}\sum_{b}m_{b}(z_{a}-z_{b})
-\frac{p^{3}_{a}}{2m^{3}_{a}}+\frac{\kappa}{4}\sum_{b}m_{b}\frac{p_{a}}{m_{a}}
r_{ab}-\frac{\kappa}{4}\sum_{b}p_{b}r_{ab}
\\
\dot{p}_{a}&=&-\frac{\partial H}{\partial z_{a}}
\nonumber \\
&=&-\frac{\kappa}{4}\sum_{b}m_{a}m_{b}\frac{\partial r_{ab}}{\partial z_{a}}
-\frac{\epsilon\kappa}{4}\sum_{b}(m_{a}p_{b}-m_{b}p_{a})-\frac{\kappa}{8}
\sum_{b}\left(m_{a}\frac{p^{2}_{b}}{m_{b}}+m_{b}\frac{p^{2}_{a}}{m_{a}}\right)
\frac{\partial r_{ab}}{\partial z_{a}}
\nonumber \\
&&+\frac{\kappa}{4}\sum_{b}p_{a}p_{b}\frac{\partial r_{ab}}{\partial z_{a}}
-\frac{1}{2}\left(\frac{\kappa}{4}\right)^{2}\sum_{b}\sum_{c}m_{a}m_{b}m_{c}
\left(\frac{\partial r_{ab}}{\partial z_{a}}r_{bc}+\frac{\partial r_{ab}}
{\partial z_{a}}r_{ac}\right)
\nonumber \\
&&-\frac{1}{2}\left(\frac{\kappa}{4}\right)^{2}\sum_{b}\sum_{c}m_{a}m_{b}m_{c}
(z_{a}-z_{b})\;\;.
\end{eqnarray}
By eliminating $p_{a}$ in the equations (68) and (69), we get the equations 
of motion to second order
\begin{eqnarray}
\lefteqn{m_{a}\ddot{z}_{a}}
\nonumber \\
&=&-\frac{\kappa}{4}\sum_{b}m_{a}m_{b}\frac{\partial r_{ab}}
{\partial z_{a}}+\frac{\kappa}{8}\sum_{b}m_{a}m_{b}\left\{4(\dot{z}_{a})^{2}
-2\dot{z}_{a}\dot{z}_{b}+(\dot{z}_{b})^{2}\right\}\frac{\partial r_{ab}}
{\partial z_{a}}
\nonumber \\
&&-\left(\frac{\kappa}{4}\right)^{2}\sum_{b}\sum_{c}m_{a}m_{b}m_{c}
\left(\frac{\partial r_{ac}}{\partial z_{a}}-\frac{\partial r_{bc}}
{\partial z_{b}}\right)r_{ab}-\frac{1}{2}\left(\frac{\kappa}{4}\right)^{2}
\sum_{b}\sum_{c}m_{a}m_{b}m_{c}\left(\frac{\partial r_{ab}}{\partial z_{a}}
r_{bc}+\frac{\partial r_{ab}}{\partial z_{a}}r_{ac}\right)
\nonumber \\
&&+\frac{1}{2}\left(\frac{\kappa}{4}\right)^{2}\sum_{b}\sum_{c}m_{a}m_{b}m_{c}
(z_{a}-z_{b})
\end{eqnarray}

We pause to show that these canonical equations of motion are equivalent
to the geodesic equation (6). To evaluate the geodesic equation, we have 
to know all components of the metric tensor, 
of which $g_{11}=\gamma$ is fixed 
and $N_{0}$ and $N_{1}$ components are determined by combining the time 
derivatives of the coordinate conditions (25) and (26) with the equations 
(13) and (16). These equations are
\begin{eqnarray}
\kappa\chi^{\prime}N_{0}+N^{\prime}_{1}=0
\\
\partial_{1}\left(\frac{1}{2}N_{0}\Psi^{\prime}+N^{\prime}_{0}\right)=0\;\;.
\end{eqnarray}
The solutions in the post-Newtonian approximation are
\begin{eqnarray}
N_{0}&=&1+\frac{\kappa}{4}\sum_{a}m_{a}r_{a}
+\frac{\epsilon\kappa}{4}\sum_{a}p_{a}(x-z_{a})
+\frac{\kappa}{8}\sum_{a}\frac{p^{2}_{a}}{m_{a}}r_{a}
+\frac{1}{4}\left(\frac{\kappa}{4}\right)^{2}
\left(\sum_{a}m_{a}r_{a}\right)^{2}
\nonumber \\
&&+\frac{1}{4}\left(\frac{\kappa}{4}\right)^{2}\left(\sum_{a}m_{a}
(x-z_{a})\right)^{2}
+\frac{1}{2}\left(\frac{\kappa}{4}\right)^{2}\sum_{a}\sum_{b}m_{a}m_{b}
r_{a}r_{ab}
\\
N_{1}&=&\frac{\epsilon\kappa}{4}\sum_{a}m_{a}(x-z_{a})+\frac{\kappa}{4}
\sum_{a}p_{a}r_{a}\;\;.
\end{eqnarray}

The boundary conditions in the previous section are incorporated in 
these solutions through (12) and (17), which serve as consistency
equations.

Since we have proved in section 2 the equivalence of the geodesic equation
(6) to a set of equations (18) and (19), we have only to evaluate the 
latter equations. Under the coordinate conditions (25) and (26), the 
equations (18) and (19) become
\begin{eqnarray}
\dot{z}_{a}&=&\frac{p_{a}}{\sqrt{p^{2}_{a}+m^{2}_{a}}}
N_{0}(z_{a})-N_{1}(z_{a})
\\
\dot{p}_{a}&=&-\sqrt{p^{2}_{a}+m^{2}_{a}}\frac{\partial N_{0}}{\partial z_{a}}
+p_{a}\frac{\partial N_{1}}{\partial z_{a}}
\end{eqnarray}
which lead to the equations (68) and (69), after the insertion of the 
solutions (73) and (74) and expanding in powers of $c^{-1}$.

The Hamiltonian (65) contains  a term proportional to
$c^{-1}$, namely, 
\[\frac{\epsilon\kappa}{8}\sum_{a}\sum_{b}(m_{a}p_{b}-m_{b}p_{a})(z_{a}-z_{b}) 
\;\;.\]
The appearance of 
such a term may seem unnatural, because in $(3+1)$ dimensions 
terms in odd powers of $c^{-1}$ , for example $c^{-5}$ terms, 
are related with gravitational radiation; yet
there exists no graviton degree of freedom in (1+1)-dimensions.
However, by a suitable redefinition of 
canonical variables 
\begin{eqnarray}
z_{a}&\longrightarrow& \tilde{z}_{a}=z_{a} 
\nonumber \\
\\
p_{a}&\longrightarrow& \tilde{p}_{a}=p_{a}-\frac{\epsilon\kappa}{4}\sum_{b}
m_{a}m_{b}(z_{a}-z_{b})\;\;
\nonumber \\
\end{eqnarray}
this term can be eliminated.
Under this redefinition the ''Poisson brackets'' among the canonical
variables are kept unchanged. The Hamiltonian expressed in terms of the 
redefined canonical variables is then
\begin{eqnarray}
H&=&\sum_{a}m_{a}+\sum_{a}\frac{\tilde{p}^{2}_{a}}{2m_{a}}
+\frac{\kappa}{8}\sum_{a}\sum_{b}m_{a}m_{b}\tilde{r}_{ab}
-\sum_{a}\frac{\tilde{p}^{4}_{a}}{8m^{3}_{a}}
+\frac{\kappa}{8}\sum_{a}\sum_{b}m_{a}\frac{\tilde{p}^{2}_{b}}{m_{b}}
\tilde{r}_{ab}
\nonumber \\
&&-\frac{\kappa}{8}\sum_{a}\sum_{b}\tilde{p}_{a}\tilde{p}_{b}\tilde{r}_{ab}
+\frac{1}{4}\left(\frac{\kappa}{4}\right)^{2}\sum_{a}\sum_{b}\sum_{c}
m_{a}m_{b}m_{c}\left\{\tilde{r}_{ab}\tilde{r}_{ac}
-(\tilde{z}_{a}-\tilde{z}_{b})(\tilde{z}_{a}-\tilde{z}_{c})\right\}\;\;.
\end{eqnarray}
It is straightforward to show that
the equations of motion derived from the Hamiltonian (79) 
are identical with those derived from the original variables.

\section{Discussion}

We have performed the canonical reduction of a set of point particles
coupled to a gravitational field in $(1+1)$ dimensions in the context
of the $R=T$ theory, and have shown how to carry out series expansions
of such equations in powers of $\kappa$ (the post-linear approximation)
and powers of $c^{-1}$ (the post-Newtonian approximation). We have
obtained explicit expressions for the Hamiltonian to order $\kappa^2$ 
and $c^{-4}$ respectively, along with an expansion of the geodesic
equation to order $c^{-4}$. 

We have considered in this reduction
the dynamical role of the auxiliary scalar
field $\Psi$. Despite the decoupling of $\Psi$ from the
classical gravity/matter field equations, the Hamiltonian 
of the system is defined in terms of $\Psi$. One of the essential points in
this analysis is the choice of the coordinate conditions, 
which we chose in the forms of (23) and (24) through the 
transformation of the generator. 
We note that our coordinate conditions are quite 
analogous to the ADM coordinate conditions in (3+1)-dimensional gravity
\cite{r6}. These coordinate conditions (frequently used for particle 
dynamics \cite{r7,r8}) are 
\begin{eqnarray}
x^{i}&=&g_{i}-\frac{1}{4\triangle}g^{T}_{\;,i}
\\
t&=&-\frac{1}{2\triangle}\pi^{ii}
\end{eqnarray}
where
\begin{eqnarray}
g_{i}&=&\frac{1}{\triangle}\left(g_{ij,j}-\frac{1}{2\triangle}
g_{jk,jki}\right)
\\
g^{T}&=&g_{ii}-\frac{1}{\triangle}g_{ij,ij}
\end{eqnarray}
and $\pi^{ij}$ is a conjugate field to $g_{ij}$, $\triangle$ being the 
Laplacian operator in three dimensions.

The differential forms of the conditions (80) and (81) are
\begin{eqnarray}
\triangle g_{ij,j}-\frac{1}{4}g_{jk,ijk}&-&\frac{1}{4}\triangle g_{jj,i}=0
\\ 
\pi^{ii}&=&0\;\;.
\end{eqnarray}
The first condition (84) relates space-derivatives of spatial
components $g_{ij}$ of the metric tensor. In $(1+1)$ dimensions 
there is only one spatial component $\gamma=g_{11}$; from (23)
the condition analogous to (84) is $\partial_{1}\gamma=0$. 
Therefore we chose the condition (25). 
Since $\pi^{ij}$ in the second condition (85) is the conjugate momenta to 
$g_{ij}$, the corresponding condition in (1+1)-dimension might
at first appear to be $\pi=0$.
However, the form of the generator (21) indicates that when the variation
$\delta h$ is related to one of the coordinate conditions, the variation of 
the conjugate $\pi$ can no longer be exploited for the other condition. The 
candidates would be $\delta\Pi$ or $\delta\Psi$. Judging from the fact that
$\pi^{ij}$ is essentially a time-derivative of $g_{ij}$ and $\Pi$ is also
a time-derivative of $\gamma$, we have no choice to take other than 
$\delta\Pi$.

We turn now to a discussion of the consistency of the full set of 
equations (12) - (19).
As discussed in previous sections, to determine the Hamiltonian for 
a system of particles, we have only to solve the constraint equations (14) 
and (15) for $\Psi$ and $\pi$. All informations on the dynamics 
of particles are included in the constraint equations. Equations (13) 
and (16) are used to determine $N_{0}$ and $N_{1}$, and equations (18) 
and (19) lead to the equations of motion. The remaining two equations (12) 
and (17) are not necessary in describing the dynamics of the
particles. However they are necessary in providing a full description
of the gravitational field. 

To check 
the consistency of the whole formalism, consider substituting the post-
linear solutions of $\Psi, \pi, N_{0}$ and $N_{1}$ into (12) and (17). 
After some calculation, it is straightforward to see 
that (12) is consistently satisfied but (17) 
is not. The reason is as follows. In the general solution to 
the constraint equation (14) 
(or (30)) $\Psi$ may contain in general an $x$-independent function 
$f(t)$. Since all other equations except (17) contain only spatial
derivatives of $\Psi$,  $f(t)$ does not contribute to either
the Hamiltonian or to the equations of motion. The function $f(t)$
is necessary only for the consistency of (17). The boundary
condition and the consideration on the surface term in the section 3 
can  be applied with $\Psi$ replaced by $(\Psi - f(t))$, the latter
function having no explicit dependence on $t$. 
After lengthy and complicated calculation of the equation (17), we 
can determine the explicit form of $f(t)$ to the order $\kappa^{2}$.

To illustrate this situation, let us consider a single static source
at the origin. In this case the fundamental equations are
\begin{eqnarray}
-\frac{1}{\kappa}\triangle\Psi&=&\kappa\pi^{2}-\frac{1}{4\kappa}
\left(\Psi^{\prime}\right)^{2}+M\delta(x)
\\
\pi^{\prime}&=&0
\end{eqnarray}
\begin{eqnarray}
\kappa\pi N_{0}+N^{\prime}_{1}&=&0
\\
\partial_{1}\left(\frac{1}{2}N_{0}\Psi^{\prime}+N^{\prime}_{0}\right)&=&0
\end{eqnarray}
\begin{eqnarray}
&&\dot{\pi}+N_{0}\left[\frac{3\kappa}{2}\pi^{2}+\frac{1}{8\kappa}
\left(\Psi^{\prime}\right)^{2}\right]+N_{1}\pi^{\prime}
+\frac{1}{2\kappa}N^{\prime}_{0}\Psi^{\prime}+N^{\prime}_{1}\pi=0
\\
&&\makebox[5em]{}\dot{\Psi}+2\kappa N_{0}\pi-N_{1}\Psi^{\prime}=0\;\;.
\end{eqnarray}
Solutions to the constraint equations (86) and (87), which satisfy the same 
boundary condition in the section 3, are
\begin{eqnarray}
\Psi&=&-\frac{\kappa M}{2}\mid x\mid + f(t)
\\
\pi&=&\chi^{\prime}=-\frac{\epsilon}{4}M \makebox[2em]{}
( \chi=-\frac{\epsilon M}{4}x )\;\;.
\end{eqnarray}
Here we included $f(t)$ in $\Psi$.

The solutions of $N_{0}$ and $N_{1}$ are
\begin{eqnarray}
N_{0}&=&e^{\frac{\kappa M}{4}\mid x\mid}
\\
N_{1}&=&\epsilon\;\frac{x}{\mid x\mid}
\left(e^{\frac{\kappa M}{4}\mid x\mid}-1\right)
\;\;.
\end{eqnarray}
Equation (90) is satisfied, whereas (91) leads to
\begin{equation}
\dot{f}(t)=\frac{\epsilon\kappa M}{2}\;\;.
\end{equation}
Then the solution of $\Psi$ becomes
\begin{equation}
\Psi=-\frac{\kappa M}{2}\mid x\mid+\frac{\epsilon\kappa M}{2}t\;\;.
\end{equation}
This shows that even for a static source the 
auxiliary field is not static. 

The Hamiltonian is
\begin{eqnarray}
H&=&-\frac{1}{\kappa}\int dx\triangle\Psi
\nonumber \\
&=&M\;\;.
\end{eqnarray}
Since
\begin{eqnarray}
g_{00}&=&-N^{2}_{0}+N^{2}_{1}
\nonumber \\
&=&1-2e^{\frac{\kappa M}{4}\mid x\mid}
\nonumber \\
&<&-1
\nonumber 
\end{eqnarray}
this solution has no event horizon.

The canonical formalism we have derived can be utilized in a number of
ways.  The most obvious of these is to obtain an explicit solution for
the $N$-body problem in $(1+1)$ dimensions (the preceding discussion 
illustrates the solution when $N=1$). In the large--$N$ limit with 
fixed proper distance between the most widely separated particles one
might expect to recover the fluid collapse problem studied in 
ref. \cite{r5}.  More generally, one could study the $N$-body problem 
where gravity is coupled to additional matter fields.  These all remain
interesting subjects for future investigation.

\vspace{1cm}
\noindent
\Large\bf Appendix  \\
\noindent
Relation between the equations of motion in the \\
coordinate conditions $\gamma=1, \Pi=0$ and those in the \\
conditions $g_{00}=-g^{-1}_{11}, g_{01}=0$
\rm\normalsize

\vspace{5mm}
One of the most popular coordinate conditions in $(1+1)$ dimensions is
the one for which the metric tensor has the form
\begin{equation}
g_{\mu\nu}=\left(
\begin{array}{cc}
-\alpha & 0 \\
0 & \alpha^{-1}
\end{array}
\right)\;\;.
\end{equation}
The difference between this condition and the condition we adopted
in this paper is that the former is independent of the
auxiliary scalar field $\Psi$ but the latter is not.
This $\Psi$-dependence of the coordinate condition played an important
role in the canonical reduction and the general expression of the 
Hamiltonian.

In this appendix we shall compare the equations of motion in 
these two coordinate conditions.
Expanding $\alpha$ and the  energy-momentum tensor (5) in  a power series 
of $c^{-1}$ we obtain
\begin{equation}
\alpha=1+2\phi-g^{(4)}_{00}+\cdot\cdot\cdot
\end{equation}
\begin{eqnarray}
T_{00}&=&\sum_{a}\frac{m_{a}}{\sqrt{1-\alpha^{-2}(\dot{z}_{a})^{2}}}
\alpha^{3/2}\delta(x-z_{a})
\nonumber \\
&=&\sum_{a}m_{a}\left\{1+3\phi+\frac{1}{2}(\dot{z}_{a})^{2}
+\cdot\cdot\cdot\right\}
\delta(x-z_{a})
\\
T_{01}&=&-\sum_{a}\frac{m_{a}\alpha^{-1/2}\dot{z}_{a}}{\sqrt{1-\alpha^{-2}
(\dot{z}_{a})^{2}}}\delta(x-z_{a})
\nonumber \\
&=&\sum_{a}m_{a}\left\{-\dot{z}_{a}+\cdot\cdot\cdot\right\}\delta(x-z_{a})
\\
T_{11}&=&\sum_{a}\frac{m_{a}\alpha^{-5/2}(\dot{z}_{a})^{2}}
{\sqrt{1-\alpha^{-2}(\dot{z}_{a})^{2}}}\delta(x-z_{a})
\nonumber \\
&=&\sum_{a}m_{a}(\dot{z}_{a})^{2}\left\{1+\cdot\cdot\cdot\right\}
\delta(x-z_{a})
\end{eqnarray}
where the number in the upper parenthesis denotes the order of $c^{-1}$
and $\phi\equiv-\frac{1}{2}g^{(2)}_{00}$.

The post-Newtonian equation of motion of the reference \cite{r5} reads
\begin{equation}
m_{1}\ddot{z}_{1}=-m_{1}\phi^{\prime}-m_{1}\psi^{\prime}
-4m_{1}\phi\phi^{\prime}+3m_{1}\dot{\phi}\dot{z}_{1}
+3m_{1}\phi^{\prime}(\dot{z}_{1})^{2}\;\;.
\end{equation}
Here $\phi$ and $\psi$ satisfy the equations
\begin{eqnarray}
\phi^{\prime\prime}&=&\frac{\kappa}{2}T^{(0)}_{00} 
\nonumber \\
&=&\frac{\kappa}{2}\sum_{a}m_{a}\delta(x-z_{a})
\\
\psi^{\prime\prime}&=&-\ddot{\phi}-2(\phi^{\prime})^{2}+\frac{\kappa}{2}
\left(T^{00(2)}-T^{11(2)}\right)
\nonumber \\
&=&-\ddot{\phi}-2(\phi^{\prime})^{2}-\frac{\kappa}{2}\sum_{a}m_{a}\left\{
\frac{\kappa}{4}\sum_{b}m_{b}r_{ab}+\frac{1}{2}(\dot{z}_{a})^{2}\right\}
\delta(x-z_{a})\;\;.
\end{eqnarray}
The solutions are
\begin{eqnarray}
\phi&=&\frac{\kappa}{4}\sum_{a}m_{a}r_{a} 
\\
\psi&=&\frac{\kappa}{8}\sum_{a}m_{a}\ddot{z}_{a}(x-z_{a})r_{a}
-\frac{\kappa}{4}\sum_{a}m_{a}(\dot{z}_{a})^{2}r_{a}
\nonumber \\
&&-\left(\frac{\kappa}{4}\right)^{2}\sum_{a}\sum_{b}m_{a}m_{b}r_{a}r_{b}
+2\left(\frac{\kappa}{4}\right)^{2}\sum_{a}\sum_{b}m_{a}m_{b}r_{ab}r_{a}
\nonumber \\
&&-\frac{\kappa^{2}}{16}\sum_{a}\sum_{b}m_{a}m_{b}r_{ab}r_{a}
-\frac{\kappa}{8}\sum_{a}m_{a}(\dot{z}_{a})^{2}r_{a}\;\;.
\end{eqnarray}
The equation of motion (104) for a system of two particles becomes
\begin{eqnarray}
m_{1}\ddot{z}_{1}&
=&-\frac{\kappa}{4}m_{1}m_{2}\frac{\partial r_{12}}{\partial z_{1}}
-\frac{\kappa^{2}}{8}m_{1}m_{2}(m_{1}+m_{2})(z_{1}-z_{2})
\nonumber \\
&&+\frac{\kappa}{8}m_{1}m_{2}\left\{6(\dot{z}_{1})^{2}-6\dot{z}_{1}\dot{z}_{2}
+3(\dot{z}_{2})^{2}\right\}\frac{\partial r_{12}}{\partial z_{1}}\;\;.
\end{eqnarray}

To investigate the relation between this equation of motion and the 
equation obtained in our canonical formalism, we consider coordinate 
transformations $x^{\mu}\longrightarrow \underline{x}^{\mu}$, which connect
two forms of the line element :
\begin{eqnarray}
ds^{2}&=&-\alpha dt^{2}+\frac{1}{\alpha}dx^{2}
\\
&=&-(N^{2}_{0}-N^{2}_{1})d\underline{t}^{2}
+2N_{1}d\underline{t}d\underline{x}+d\underline{x}^{2}\;\;,
\end{eqnarray}
The explicit form of the transformations is given by
\begin{eqnarray}
t&=&\underline{t}-\frac{\epsilon\kappa}{8}\sum_{a}m_{a}(\underline{x}
-\underline{z}_{a})^{2}+\cdot\cdot\cdot
\\
x&=&\underline{x}+\frac{\kappa}{8}\sum_{a}m_{a}(\underline{x}
-\underline{z_{a}})
\mid\underline{x}-\underline{z_{a}}\mid+\cdot\cdot\cdot\;\;.
\end{eqnarray}

For a system of two particles the above transformation (113) leads to
\begin{eqnarray}
z_{1}&=&\underline{z}_{1}+\frac{\kappa}{8}m_{2}(\underline{z}_{1}
-\underline{z}_{2})
\mid\underline{z}_{1}-\underline{z}_{2}\mid+\cdot\cdot\cdot
\nonumber \\
z_{2}&=&\underline{z}_{2}-\frac{\kappa}{8}m_{1}(\underline{z}_{1}
-\underline{z}_{2})
\mid\underline{z}_{1}-\underline{z}_{2}\mid+\cdot\cdot\cdot
\nonumber \\
z_{1}-z_{2}&=&(\underline{z}_{1}-\underline{z}_{2})\left\{1+\frac{\kappa}{8}
(m_{1}+m_{2})\mid\underline{z}_{1}-\underline{z}_{2}\mid
+\cdot\cdot\cdot\right\}
\nonumber \\
r_{12}&=&\underline{r}_{12}\left\{1+\frac{\kappa}{8}(m_{1}+m_{2})
\underline{r}_{12}+\cdot\cdot\cdot\right\}\;\;.
\nonumber 
\end{eqnarray}
Since the second term of the right hand side of the transformation (112)
contributes to $\underline{t}$ as a correction of the order $c^{-3}$,
in the post-Newtonian approximation we can set $t=\underline{t}$.
Then we have
\begin{eqnarray}
\dot{z}_{1}&=&\dot{\underline{z}}_{1}+\frac{\kappa}{8}m_{2}\left\{
(\dot{\underline{z}}_{1}-\dot{\underline{z}}_{2})\mid\underline{z}_{1}
-\underline{z}_{2}\mid
+(\underline{z}_{1}-\underline{z}_{2})(\dot{\underline{z}}_{1}
-\dot{\underline{z}}_{2})
\frac{\partial\underline{r}_{12}}{\partial\underline{z}_{1}}\right\}
+\cdot\cdot\cdot
\nonumber \\
&=&\dot{\underline{z}}_{1}+\frac{\kappa}{4}m_{2}(\dot{\underline{z}}_{1}
-\dot{\underline{z}}_{2})\mid\underline{z}_{1}-\underline{z}_{2}\mid
+\cdot\cdot\cdot
\nonumber \\
\ddot{z}_{1}&=&\ddot{\underline{z}}_{1}+\frac{\kappa}{4}m_{2}\left\{
(\ddot{\underline{z}}_{1}-\ddot{\underline{z}}_{2})\mid\underline{z}_{1}
-\underline{z}_{2}\mid
+(\dot{\underline{z}}_{1}-\dot{\underline{z}}_{2})^{2}
\frac{\partial\underline{r}_{12}}
{\partial\underline{z}_{1}}\right\}+\cdot\cdot\cdot
\nonumber 
\end{eqnarray}
\begin{eqnarray}
\frac{\partial r_{12}}{\partial z_{1}}&=&\frac{\partial}{\partial z_{1}}
\left\{\underline{r}_{12}\left[1+\frac{\kappa}{8}(m_{1}+m_{2})
\underline{r}_{12}+\cdot\cdot\cdot\right]\right\}
\nonumber \\
&=&\left\{1-\frac{\kappa}{4}(m_{1}+m_{2})\underline{r}_{12}
+\cdot\cdot\cdot\right\}
\left\{\frac{\partial\underline{r}_{12}}{\partial\underline{z}_{1}}
+\frac{\kappa}{4}(m_{1}+m_{2})\frac{\partial\underline{r}_{12}}
{\partial\underline{z}_{1}}\underline{r}_{12}+\cdot\cdot\cdot\right\}
\nonumber \\
&=&\frac{\partial\underline{r}_{12}}{\partial\underline{z}_{1}}
+\cal{O}\mit(\kappa^{2})\;\;.
\nonumber
\end{eqnarray}

By the use of these relations the equation of motion (109) is transformed to
\begin{equation}
m_{1}\ddot{\underline{z}}_{1}=-\frac{\kappa}{4}m_{1}m_{2}
\frac{\partial\underline{r}_{12}}{\partial\underline{z}_{1}}+\frac{\kappa}{8}
\left\{4(\dot{\underline{z}}_{1})^{2}-2\dot{\underline{z}}_{1}
\dot{\underline{z}}_{2}
+(\dot{\underline{z}}_{2})^{2}\right\}\frac{\partial\underline{r}_{12}}
{\partial\underline{z}_{1}}-\frac{\kappa^{2}}{16}m_{1}m_{2}(m_{1}+m_{2})
(\underline{z}_{1}-\underline{z}_{2})
\\
\end{equation}
This equation is identical with the equation of motion (70) we obtained in
the canonical formalism.

\section*{Acknowledgements} This work was supported in part by the
Natural Sciences and Engineering Research Council of Canada.


\vspace{3cm}
\begin{thebibliography}{1}

\bibitem{r1}
Teitelboim C
1984
{\it Quantum Theory of Gravity}
ed S Christensen (Bristol: Hilger) p 327

Jackiw R
1984
{\it Quantum Theory of Gravity}
ed S Christensen (Bristol: Hilger) p 403; 1985 
{\it Nucl. Phys} B \bf 252\rm \hspace{2mm}343

\bibitem{r2}
Marnelius R 
1983
{\it Nucl. Phys.}
\bf B 211\rm \hspace{2mm}14

Torre C G 
1989
{\it Phys. Rev.}
\bf D 40\rm \hspace{2mm}2588

Mann R B, Shiekh A and Tarasov L
1990
{\it Nucl. Phys.}
\bf B 341\rm \hspace{2mm}134  

\bibitem{HS92} Harvey J and Strominger A
1992
 ``Quantum Aspects of Black Holes,'' {\tt hep-th/9209055} 


\bibitem{BanksMann}
Banks T and O' Loughlin M
1991
{\it Nucl. Phys.} {\bf B362 } \hspace{2mm} 649 

Mann RB
1993
{\it Phys. Rev.} 
{\bf D47} \hspace{2mm} 4438

\bibitem{LMK}
Louis-Martinez D and Kunstatter G
1995
{\it Phys. Rev.} {\bf D52} \hspace{2mm} 3494


\bibitem{2djol}
Creighton JDE and Mann RB
1995
``Thermodynamics of Dilatonic Black Holes in N-dimensions''
{\tt gr-qc/9511012} 


\bibitem{jchan}
Chan SFJ and Mann RB
1995
{\it Class. Quant. Grav. } 
{\bf 12} \hspace{2mm} 351

\bibitem{RST}
Russo J, Susskind L and Thorlacius L
1993
{\it Phys. Rev.} {\bf D47} 533

\bibitem{r3}
Mann R B
1991
{\it Found. Phys. Lett.}
\bf 4\rm \hspace{2mm}425

Mann R B
1992
{\it Gen. Rel. Grav.}
\bf 24\rm \hspace{2mm}433


\bibitem{r4}
Mann R B, Morsink S M, Sikkema A E and Steele T G
1991
{\it Phys. Rev.}
\bf D 43\rm \hspace{2mm}3948

Morsink S and Mann RB
1991
{\it Class. Quantum Grav.} \bf 8\rm \hspace{2mm} 2257

Mann RB and Steele TG
1992
{\it Class. Quantum Grav.} \bf 9\rm \hspace{2mm} 475


\bibitem{r5}
Sikkema A E and Mann R B 
1991 
{\it Class. Quantum Grav.} 
\bf 8\rm \hspace{2mm}219

Christensen JD and Mann RB
1992
{\it Class. Quantum Grav.} \bf 9\rm \hspace{2mm} 1769

Chan KCK and Mann RB 
1993 
{\it Class. Quantum Grav.} \bf 10\rm \hspace{2mm} 913


\bibitem{rtquant}
Mann RB
1994
{\it Nucl. Phys.} {\bf B418} \hspace{2mm} 231 

\bibitem{2dross}
Mann RB and Ross SF 
1993
{\it Class. Quantum Grav.} \bf 10\rm \hspace{2mm} 1405



\bibitem{r6}
Arnowitt R, Deser S and Misner C W
1962
{\it Gravitation: An Introduction to Current Research, Chap.7}
(New York: Wiley) \hspace{2mm}
This is a review work of a series of their papers, which are listed in the
references of this paper. 

\bibitem{r7}
Kimura T
1961
{\it Prog. Theor. Phys.}
\bf 26\rm \hspace{2mm}157

\bibitem{r8}
Ohta T and Kimura T
1989
{\it Classical and Quantum Gravity, Chap.6 (in Japanese)}
(Tokyo: MacGrawhill)

Ohta T, Okamura H, Kimura T and Hiida K
1974
{\it Prog. Theor. Phys}
\bf 51\rm \hspace{2mm}1598


\bibitem{r9}
Arnowitt R, Deser S and Misner C W
1960
{\it J. Math. Phys.}
\bf 1\rm \hspace{2mm}434


\bibitem{r10}
Arnowitt R, Deser S and Misner C W
1960
{\it Phys. Rev.}
\bf 117\rm \hspace{2mm}1595


\end{thebibliography}
\end{document}